\documentclass[twocolumn,superscriptaddress,floatfix,showpacs, prb,10pt]{revtex4-1}
\usepackage[english]{babel}
\usepackage{graphicx}
\usepackage{amsmath}
\usepackage{amssymb}
\usepackage{bm}
\usepackage{color}

\begin{document}

\title{Momentum-space structure of surface states in a topological semimetal with a nexus point of Dirac lines}

\author{T. Hyart}
\affiliation{Department of Physics and Nanoscience Center, University of Jyv\"askyl\"a, P.O. Box 35 (YFL), FI-40014 University of Jyv\"askyl\"a, Finland}

\author{T.~T.~Heikkil\"a}
\affiliation{Department of Physics and Nanoscience Center, University of Jyv\"askyl\"a, P.O. Box 35 (YFL), FI-40014 University of Jyv\"askyl\"a, Finland}


\begin{abstract}
Three-dimensional topological semimetals come in different variants, either containing Weyl points or  Dirac lines. Here we describe a more complicated momentum-space topological defect where several separate Dirac lines connect with each other, forming a momentum-space equivalent of the real-space nexus considered before for Helium-3. Close to the nexus the Dirac lines exhibit a transition from type I to type II lines. We consider a general model of stacked honeycomb lattices with the symmetry of Bernal (AB) stacked graphite and show that the structural mirror symmetries in such systems protect the presence of the Dirac lines, and also naturally lead to the formation of the nexus. By the bulk-boundary correspondence of topological media, the presence of Dirac lines lead to the formation of drumhead surface states at the side surfaces of the system. We calculate the surface state spectrum, and especially illustrate the effect of the nexus on these states. 
\end{abstract}

\maketitle

\section{Introduction} The study of momentum space topological defects is an important topic in modern condensed matter physics \cite{Volovik-book, HaKa10, Zhang-review}. 
Topological materials are characterised by the existence of bulk topological invariants and protected surface states. The fully gapped topological phases have been classified in terms of the existence of various antiunitary symmetries---e.g.~time-reversal and particle-hole symmetries \cite{Schny+08, Kitaev-class}---and also the importance of the unitary symmetries is understood \cite{Fu11, Fang12, Slager13, Zhang13, Chiu13, Shiozaki14, Kimme16}. 
Recently, there have been attempts to classify also the gapless topological phases \cite{Horava,Zhao13, Matsuura-review, Chiu14, Yang14, Gao15, Zhao16, Yang16}.
In an ordinary $d$-dimensional metal a $d-1$-dimensional Fermi surface separates the filled and empty states. On the other hand, topological semimetals and nodal superconductors exhibit  lower dimensional Fermi lines or Fermi points, whose stability is guaranteed by certain symmetries and a nontrivial topology of the wave functions. The simplest examples of gapless topological phases are graphene \cite{Ryu02} and $d$-wave superconductors \cite{CuTheory, Nakada96,Fujita96} supporting topological flat bands at the edges. Three-dimensional topological semimetals with Dirac points \cite{Liu14, Neupane14, Xu15}, Weyl points \cite{Wan11, Weng15a,  Huang15a, Ruan16, Xu15b, Lv15, Lv15b,Yang15, Xu15c, Inoue16} and Dirac lines \cite{Heikkila-flat-bands, Mullen15, Kim15, Yu15, Xie15, Weng15, Bian16, Zeng15, Bian15, Schoop15, Chan15, Wu16,Chang16, Weng16, Henni16,Li16} have been theoretically predicted and experimentally observed. Moreover, nodal topological band structures with topologically protected surface states can arise in superconductors with unconventional pairing symmetries \cite{Volovik87, Volovik-book, Tanaka10, SchnyderRyu11, Brydon11, Sato11, KYada2011,Schnyder12, TanakaSato12, Lee12, Hyart14, Schnyder15, Volovik-polar, Autti15}.

The possible topological defects occurring in gapless materials are not limited to these options.
The next step in the increasing complexity of the momentum space topological defects is to consider the topology of the systems containing multiple  Dirac lines \cite{nodalchain, nexus}. These Dirac lines can meet somewhere in the momentum space giving rise to an exceptional point, which can be called as nexus \cite{nexus}. In real space nexuses can appear when vortex lines meet and such kinds of topological defects have been considered  in the contexts of superfluid Helium-3 \cite{Volovik-book} and as a possible mechanism for confinement in the Georgi-Glashow model in particle physics \cite{Cornwall99}. 
Recently, the idea that a momentum space nexus may occur in topological semimetals was put forward in Ref.~\onlinecite{nexus} in the context of Bernally (AB) stacked graphite, where the existence and merging of multiple band contact lines is theoretically well-established \cite{McClure57, Mikitik06,Mikitik08}.

\begin{figure}
  \includegraphics[width=0.99\columnwidth]{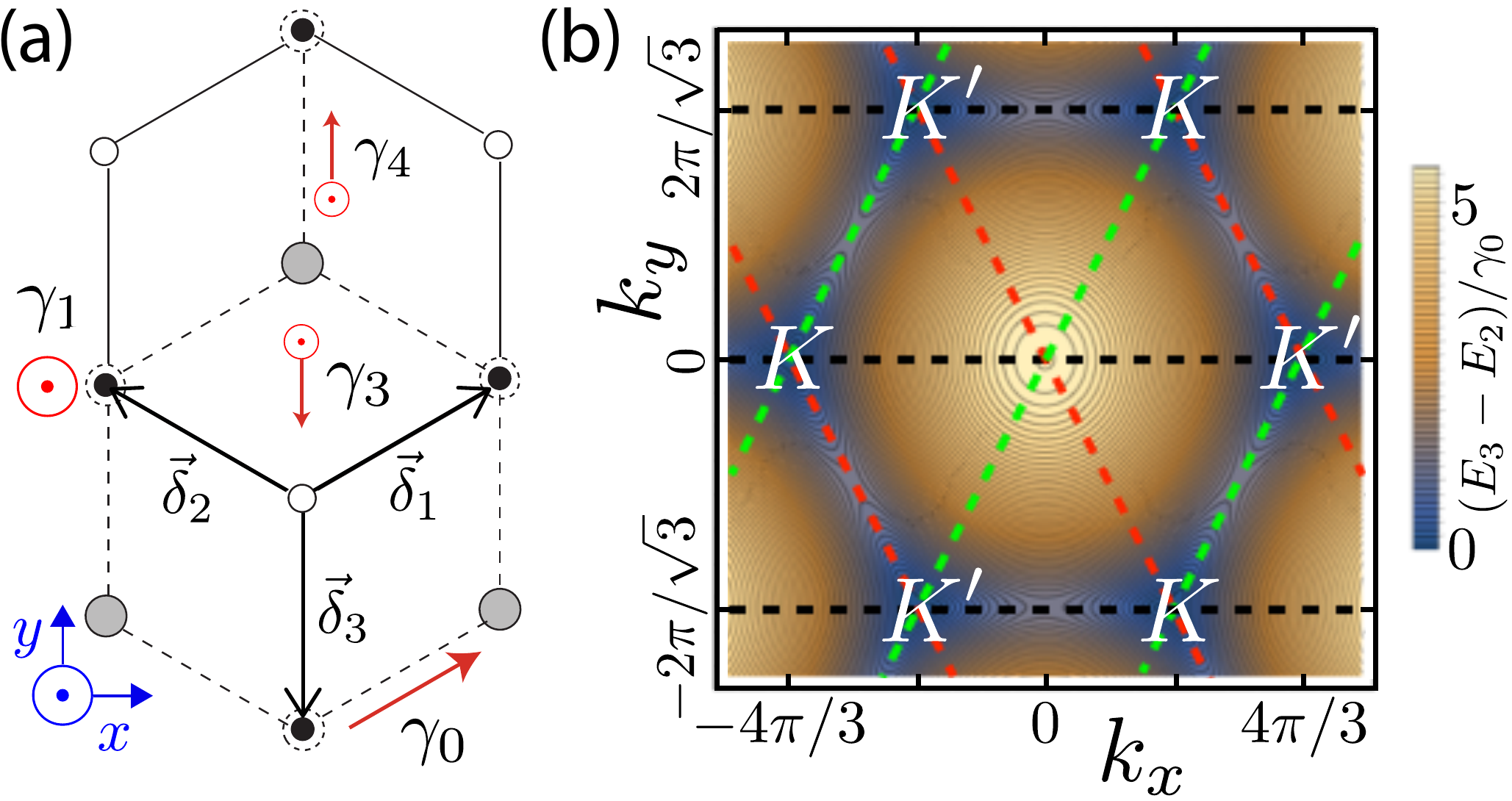}
  \caption{\label{fig:schematic}  (a) Unit cell of Bernally stacked honeycomb lattice: two layers (solid and dashed lines), each consisting of two sublattices (filled and unfilled circles). The hopping terms $\gamma_i$ are depicted (hopping  between the layers is indicated with $\odot$). (b) Momentum-dependent energy gap between the two bands closest to the Fermi level $E_3(\mathbf{k})-E_2(\mathbf{k})$ as a function of $k_x$ and $k_y$ for $k_z=0$. There are band crossings in the vicinity of $K$ and $K'$ points in the $(k_x, k_y)$-plane (Fig.~\ref{fig:Diraclines}). As a function of $k_z$ these band crossings form nodal Dirac lines, so that for each momentum along the line the energy spectrum with respect to the transverse momenta is conical. The band crossings always appear within the  mirror planes directed along the $k_z$ direction and the dashed lines within the $(k_x, k_y)$-plane, and are protected by structural mirror symmetries. The mirror planes share common lines in the momentum space directed along $k_z$-direction exactly at $K$ and $K'$ in the $(k_x, k_y)$-plane. The Dirac lines can merge at a specific $k_z$ within this line, leading to an appearance of a nexus. 
}
\end{figure}

In this paper we study the properties of the nexus semimetal phase proposed in Refs.~\onlinecite{nexus, McClure57, Mikitik06,Mikitik08}. We identify the key properties of the model that allow stabilizing the nexus in the momentum space:  reflection symmetries and the existence of more than one electron- or hole-like bands  close to the Fermi energy. The reflection symmetries of the structure specify mirror planes in the momentum space where the Hamiltonian is block-diagonal, and allow stabilizing band crossings where the bands have different eigenvalues of the mirror symmetry operator \cite{Chiu14, Chan15}. The requirement for stabilizing the nexus is to have several of those mirror planes sharing a common line in the momentum space,  so that stable Dirac lines can merge at a specific point within this line.
We show that in the nexus semimetal phase proposed in Refs.~\onlinecite{nexus, McClure57, Mikitik06,Mikitik08} the nexus indeed defines a point-like topological defect in the momentum space where three bands are degenerate at the same momentum, and the appearance of this type of an exceptional point distinguishes the nexus semimetal phase from a semimetal containing (multiple) Dirac lines that do not merge in the momentum space. By the bulk-boundary correspondence of topological media, the presence of Dirac lines lead to the formation of surface states. We  show that the number of surface states as a function of the momentum components parallel to the surface displays a fine-structure associated with the projected Dirac lines. Far away from the nexus the surface state dispersions take a form of a drumhead that is bounded by the projected crossing points of the electron- and hole-like bands \cite{Chiu14,Chan15}. In the vicinity of the nexus there occurs a crossover to a different type of behavior, where one of the surface bands connects two electron-like bands to each other, and the other surface band connects two hole-like bands. In the crossover regime the surface states hybridize with the bulk states so that they connect bulk band edges to each other instead of being bounded by the projected Dirac lines. 

\section{Model and symmetries} 
We consider Bernally stacked honeycomb lattices [Fig.~\ref{fig:schematic}(a)] such as Bernal graphite \cite{nexus, McClure57, Mikitik06,Mikitik08}. The tight-binding model for such kind of three-dimensional system [in the layer $\otimes$ sublattice space described in Fig.~\ref{fig:schematic}(a)] can be written as 
\begin{widetext}
\begin{equation}
H(k_x, k_y, k_z)= \begin{pmatrix}
  \Delta      & - \gamma_0 f(k_x,k_y) &  2 \gamma_4 \Gamma(k_z)  f^*(k_x,k_y)      & -2 \gamma_1 \Gamma(k_z) \\
- \gamma_0 f^*(k_x,k_y)    & 0 & 2 \gamma_3 \Gamma(k_z)  f(k_x,k_y)     &  2 \gamma_4 \Gamma(k_z)  f^*(k_x,k_y)	   \\
 2 \gamma_4 \Gamma(k_z)  f(k_x,k_y)           & 2 \gamma_3 \Gamma(k_z)  f^*(k_x,k_y)         & 0  & - \gamma_0 f(k_x,k_y)  \\
   -2 \gamma_1 \Gamma(k_z)            & 2 \gamma_4 \Gamma(k_z)  f(k_x,k_y)       &- \gamma_0 f^*(k_x,k_y)  & \Delta  
\end{pmatrix}. \label{full-H}
\end{equation}
\end{widetext}
The different hopping parameters $\gamma_i$  are illustrated in Fig.~\ref{fig:schematic}(a). We have neglected further neigbor hoppings, as they do not change the qualitative conclusions if they are small and preserve the structural symmetry. Additionally, there exists a parameter $\Delta$, denoting a locally broken A-B sublattice symmetry but still preserving the global A-B symmetry. This term is allowed by the symmetries of the structure, since one of the sublattices in each layer has an atom on top of it so that it has a different environment than the other sublattice. The structure factors arising from the Fourier transform are $f(k_x,k_y)=\sum e^{i \vec{\delta}_i \cdot (k_x,k_y)}$ and $\Gamma(k_z)=\cos(k_z)$. The nearest neighbor vectors $\vec{\delta}_i$ [Fig.~\ref{fig:schematic}(a)] are normalized so that the vectors connecting neighboring unit cells have unit length. In $z$-direction we use the spacing between the layers as the unit length so that $-\pi/2 \leq k_z \leq \pi/2$. 

The most important symmetries of the model (see App.~\ref{AppA} for more details)  are a mirror symmetry
$$H(k_x, k_y, k_z)=\tau_x  \sigma_xH(k_x, -k_y, k_z)\tau_x  \sigma_x$$ and a three-fold rotational symmetry  
$$H(k_x, k_y, k_z)=H(\bar{k}_x, \bar{k}_y, k_z)=H(\tilde{k}_x, \tilde{k}_y, k_z),$$
where ($\bar{k}_{x}$, $\bar{k}_{y}$) and ($\tilde{k}_{x}$, $\tilde{k}_{y}$) are the momentum coordinates after a rotation by $\pm 2\pi/3$ around the $z$-axis. There are similar mirror symmetries also with respect to ($\bar{k}_{x}$, $\bar{k}_{y}$) and ($\tilde{k}_{x}$, $\tilde{k}_{y}$). In a special case $\Delta=\gamma_4=0$ the system supports an  accidental chiral symmetry $C H(\mathbf{k}) C=- H(\mathbf{k})$, where $C=\tau_0 \sigma_z$. In graphite, $\Delta,\gamma_4 \ll \gamma_0,\gamma_1,\gamma_3$, so that the chiral symmetry is valid as a good approximation.

There are special planes going through the $\Gamma$ point and at the boundary of the Brillouin zone which are mapped  back to themselves in the mirror symmetries (up to a reciprocal lattice vector) [Fig.~\ref{fig:schematic}(b)].  The relevant three planes around the $K$ point are directed along the $k_z$-direction and $k_y=2 \pi/\sqrt{3}$,  $\bar{k}_y=-2 \pi/\sqrt{3}$ and $\tilde{k}_y=0$ within the $(k_x, k_y)$-plane \cite{footnote}. Within these mirror planes the mirror symmetries give rise to  symmetries commuting with the Hamiltonian at fixed momentum, e.g.~ 
\begin{align*}
S^\dag H(k_x, 2 \pi/\sqrt{3} ,k_z) S&=H(k_x, 2 \pi/\sqrt{3} ,k_z).
\end{align*}
The symmetry operators in different coordinates are  $S=U \tau_x \sigma_x U^\dag$, $\bar{S}=U^\dag \tau_x \sigma_x U$, $\tilde{S}=\tau_x \sigma_x$, where $U={\rm diag}(e^{-i 2\pi/3}, 1, e^{i 2\pi/3}, e^{-i 2\pi/3})$ (see App.~\ref{AppA}), and all of them are  simultaneously valid within the line along $k_z$-direction at the $K$ point: $K=(2 \pi/3, 2\pi/\sqrt{3})$, $\bar{K}=(2 \pi/3, -2\pi/\sqrt{3})$ and $\tilde{K}=(-4\pi/3, 0)$ in different coordinates. 
 
In addition to the symmetries we assume a hierarchy of couplings $|\gamma_0| \gg |\gamma_1| > |\gamma_3|, |\gamma_4|, |\Delta|$. In particular, this fixes the overall behavior of the different bands so that two of the bands are electron-like (hole-like), bending upwards (downwards) in energy when moving away from the $K$ point.  Some of the details discussed below depend on the relative signs of the couplings. In the main text we consider all couplings to be positive [in the convention defined by the Hamiltonian (\ref{full-H})]  and we illustrate the main effects of other choices of signs in App.~\ref{othersigns}. When not otherwise stated, we choose in the figures $\gamma_1=0.3 \gamma_0$, $\gamma_3=\Delta=0.1 \gamma_0$ and $\gamma_4=0.05 \gamma_0$. These are close to the values \cite{graphenereview} often used for Bernal graphite, except that we use somewhat larger $\Delta$ and $\gamma_4$ to better illustrate the properties of the nexus.

\begin{figure}
 \includegraphics[width=1\columnwidth]{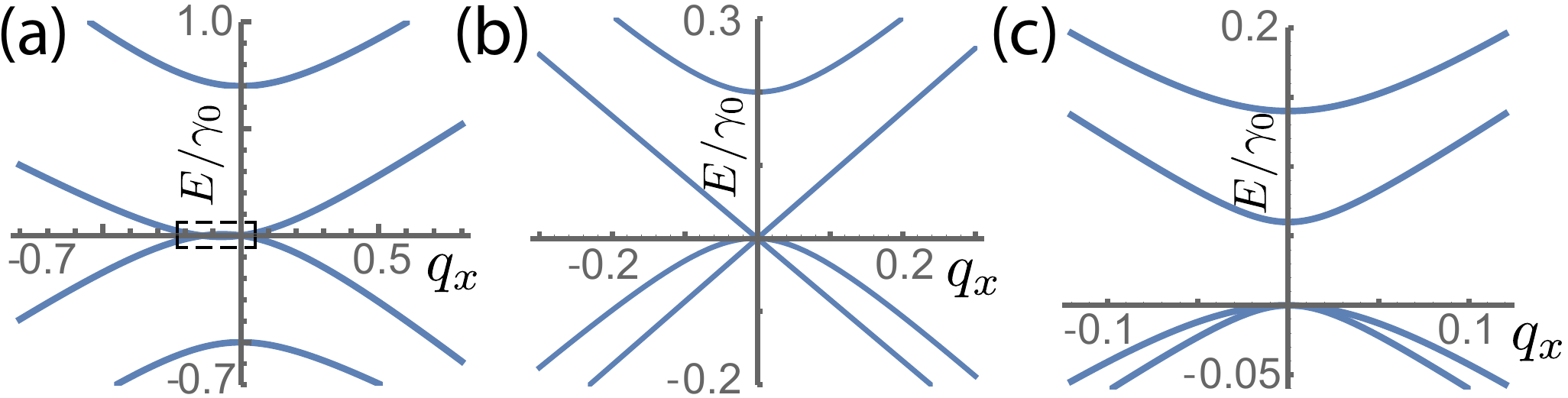}
  \caption{\label{fig:nexus} Energy-momentum dispersions for bulk bands around the $K$ point for (a) $\Gamma(k_z)= 1$, (b) $\Gamma(k_z)= \Delta/(2 \gamma_1)$ and (c) $\Gamma(k_z)= \Delta/(2 \gamma_1)-0.1$. For small $k_z$ [$\Gamma(k_z) > \Delta/(2 \gamma_1)$] electron and hole bands are degenerate at the $K$ point (a), whereas for large $k_z$ [$\Gamma(k_z) < \Delta/(2 \gamma_1)$] two hole bands are degenerate at the $K$ point and there is a gap between the electron and hole bands (c). Because the mirror symmetries guarantee that at least two bands must always be degenerate at the $K$ point, there necessarily exists a value of $k_z$ where three bands are simultaneously degenerate (b). 
}
\end{figure}

\begin{figure}
  \includegraphics[width=1\columnwidth]{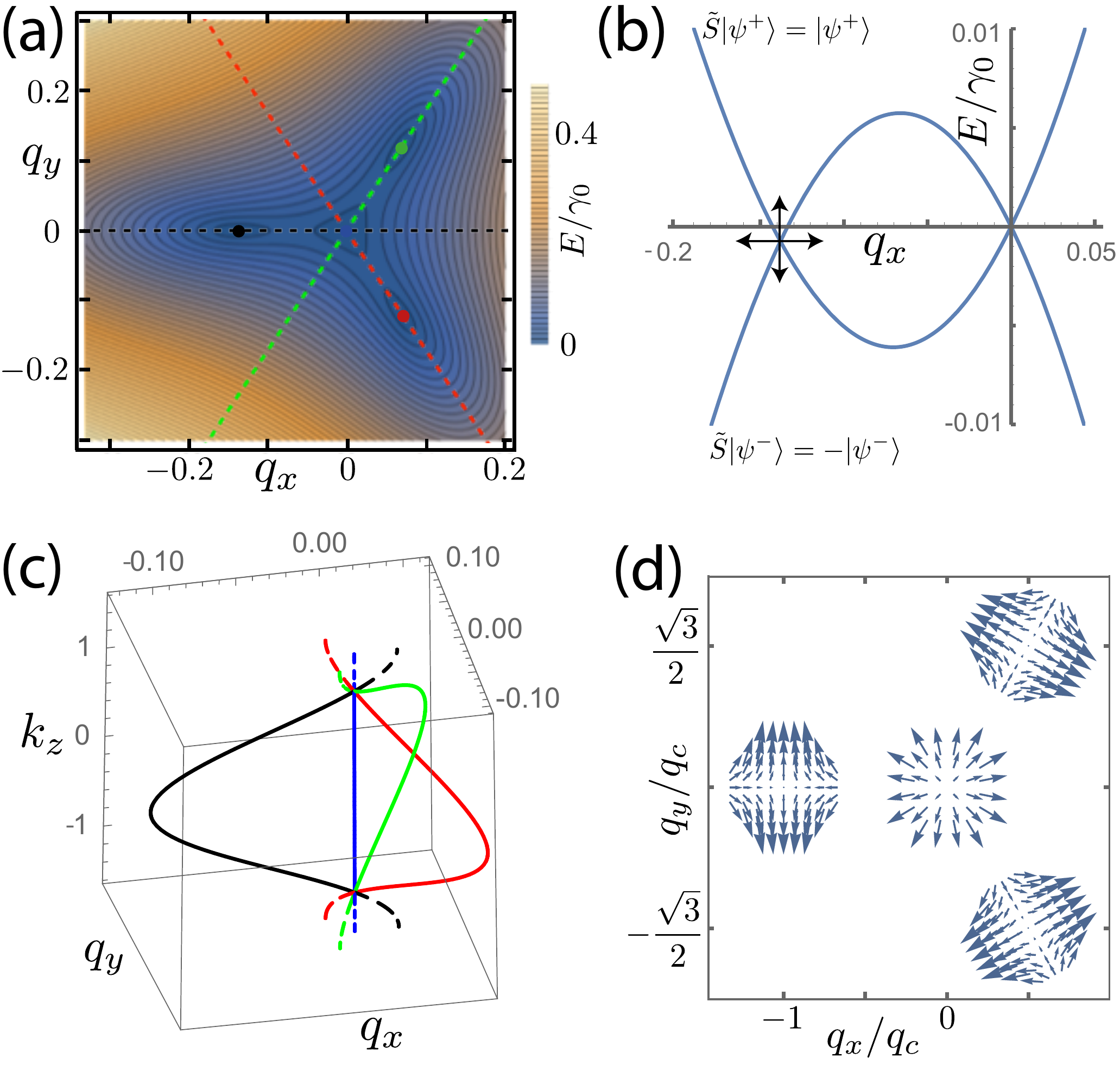}
  \caption{\label{fig:Diraclines} (a) $E_3(\mathbf{k})-E_2(\mathbf{k})$ around the $K$ point  for $k_z=0$.  There are four band crossings at $\mathbf{q}=\mathbf{0}$, $\mathbf{q}=q_c (-1,0)$ and $\mathbf{q}=q_c (1/2, \pm \sqrt{3}/2)$. As a function of $k_z$ they form Dirac lines. (b) Bulk dispersions showing the crossings at $\mathbf{q}=\mathbf{0}$ and $\mathbf{q}=q_c (-1,0)$ [boxed region in Fig.~\ref{fig:nexus}(a)]. Because the bands correspond to different mirror eigenvalues  it is only possible to move the crossing in energy and momentum (indicated with arrows), but it is not possible to remove it unless the perturbation breaks the mirror symmetry or causes a merging of several Dirac lines. (c) The Dirac lines merge at the nexus point $\mathbf{q}=\mathbf{0}$ and $\Gamma(k_z)= \Delta/(2 \gamma_1)$. For  $k_z$ above the nexus [$\Gamma(k_z) < \Delta/(2 \gamma_1)$] there is a gap between electron and hole bands [Fig.~\ref{fig:nexus}(c)], but there still exist band crossings between the two hole bands (dashed lines). (d) The low energy theories around the band crossings are described by Hamiltonians (\ref{H0Hc}). The corresponding pseudomagnetic field $\vec{h}(\delta \mathbf{q})$ is a two-component vector (represented with arrows) that forms a vortex line along the Dirac line. Therefore, the Berry phase around the Dirac line is $\pm \pi$.  In (c) we have used $\Delta=0.25 \gamma_0$ to improve the visibility of the nexus.}
\end{figure}

\section{Bulk properties} 
In the following we study the bulk spectrum around the $K$-point [$(\tilde{k}_{x}, \tilde{k}_{y})=\tilde{K}+(q_{x}, q_{y})$]  \cite{footnote}.
 Because  $\tilde{S} S={\rm diag}(1, e^{i 2\pi/3}, e^{-i 2\pi/3}, 1)$ and $\tilde{S}$ commute with the Hamiltonian, two bands must always be degenerate at the $K$ point for all values of $k_z$. This degeneracy also follows from the tight-binding Hamiltonian (\ref{full-H}) but it is protected by the existence of the mirror symmetries, not by the explicit form of the Hamiltonian. However, there are two topologically distinct options on how two bands can be degenerate within this model as illustrated in Fig.~\ref{fig:nexus}. For small $k_z$ 
electron and hole bands are degenerate at the $K$ point [Fig.~\ref{fig:nexus}(a)], whereas for large $k_z$  
two hole bands are degenerate at the $K$ point and there is an energy gap between the electron and hole bands  [Fig.~\ref{fig:nexus}(c)]. Because the mirror symmetries guarantee that at least two bands are always degenerate at the $K$ point, there must exist a value of $k_z$ [$\Gamma(k_z)= \Delta/(2 \gamma_1)$]
where three bands are simultaneously degenerate [Fig.~\ref{fig:nexus}(b)]. We call this exceptional point nexus because it is also a merging point of several Dirac lines (see below). The existence of the exceptional point where three bands are simultaneously degenerate is protected by the fact that situations (a) and (c) are topologically distinct in the presence of the mirror symmetries. 

In addition to the band contact line at the $K$ point, three other band crossings appear  \cite{nexus, McClure57, Mikitik06,Mikitik08} at energy 
$E_c$  within the three mirror planes at  $\mathbf{q}\equiv (q_x,q_y)=q_c (-1,0)$ and $\mathbf{q}=q_c (1/2, \pm \sqrt{3}/2)$ [Fig.~\ref{fig:Diraclines}(a)], where (see App.~\ref{AppA})
\begin{equation*}
E_c(k_z), q_c(k_z) \propto [ 4 \gamma_1^2 \Gamma^2 (k_z) -\Delta ^2].
\end{equation*}
 The two bands crossing at $\mathbf{q}=q_c (-1,0)$ correspond to different mirror eigenvalues  $\tilde{S} | \psi^\pm \rangle=\pm | \psi^\pm \rangle$. Therefore  $\langle \psi^- | H_1 | \psi^+ \rangle= \langle \psi^- | \tilde{S}^\dag H_1 \tilde{S} | \psi^+ \rangle=-\langle \psi^- | H_1 | \psi^+ \rangle$, so that an arbitrary perturbation $H_1$ obeying the mirror symmetry $\tilde{S}^\dag H_1 \tilde{S} =H_1$ cannot open a gap at the crossing [Fig.~\ref{fig:Diraclines}(b)].  The  number of states with mirror eigenvalue $+1$ and energy below the crossing  defines a mirror index, which is a $\mathbb{Z}$ topological invariant and has different values on the opposite sides of the crossing \cite{Chiu14, Chan15}.

The stable Dirac lines meet and merge at the nexus point $\mathbf{q}=\mathbf{0}$ and $\Gamma(k_z)= \Delta/(2 \gamma_1)$, which is shared by all the mirror planes [Fig.~\ref{fig:Diraclines}(c)]. This merging occurs at the same point where the three bands are simultaneously degenerate [Fig.~\ref{fig:nexus}(b)] \cite{footnotec}. For large $k_z$ above the nexus [$\Gamma(k_z) < \Delta/(2 \gamma_1)$] there is a gap between electron and hole bands [Fig.~\ref{fig:nexus}(c)], but there still exists band crossings between the two hole bands [dashed lines in Fig.~\ref{fig:Diraclines}(c)]. 

Around the crossings at $\mathbf{q}=\mathbf{0}$ and  $\mathbf{q}=q_c(-1, 0)$  the projected Hamiltonians in the basis of eigenvectors corresponding to eigenvalues $+1$ and $-1$ of the mirror symmetry operator at the crossings are \cite{footnoteb}
\begin{eqnarray}
H_0 &\approx& \sqrt{3} \gamma_3 \Gamma(k_z) \delta q_x \sigma_z+\sqrt{3} \gamma_3 \Gamma(k_z) \delta q_y \sigma_y, \nonumber \\
H_c&\approx&\bigg(E_c(k_z)+\frac{\sqrt{3} \gamma_3 \big[8 \gamma_1 \gamma_4 \Gamma^2(k_z)  -\Delta \gamma_0 \big] }{\gamma_1 \gamma_0} \delta q_x\bigg)  \sigma_0 \nonumber \\&&- \sqrt{3} \gamma_3 \Gamma(k_z) \delta q_x \sigma_z+3 \sqrt{3} \gamma_3 \Gamma(k_z) \delta q_y \sigma_y. \label{H0Hc}
\end{eqnarray}
The latter describes a tilted Dirac cone for each value of $k_z$, and it becomes overtilted at 
$\Gamma(k_z^*) \approx \frac{\Delta}{\gamma_1}$.
For $k_z > k_z^*$, the Dirac line can thus be called type II Dirac line \cite{Volovik_typeII_main} (see App.~\ref{AppA}).
These Hamiltonians are locally of the form 
$$H_{0 [c]} = \xi_{0 [c]}(k_z,\delta\mathbf{q}) \sigma_0 + \vec{h}_{0[c]}(k_z,\delta\mathbf{q}) \cdot \vec{\sigma},$$
where the direction of the pseudomagnetic field $\vec{h}_{0[c]}(\delta \mathbf{q})$ rotates by $2\pi$ when $\delta \mathbf{q}$ goes around the band crossing point [Fig.~\ref{fig:Diraclines}(d)]. Therefore, the wave vector would obtain a Berry phase $\pm \pi$ if taken around such a path. Such Berry phases are typically associated with the appearance of surface states \cite{Chan15}. However, the Berry phase is only defined modulo $2\pi$, and as a result this argument can  be used to explain the existence of only a single protected surface state at a given surface momentum.

\section{Surface state spectrum} 

\begin{figure}
  \includegraphics[width=1\columnwidth]{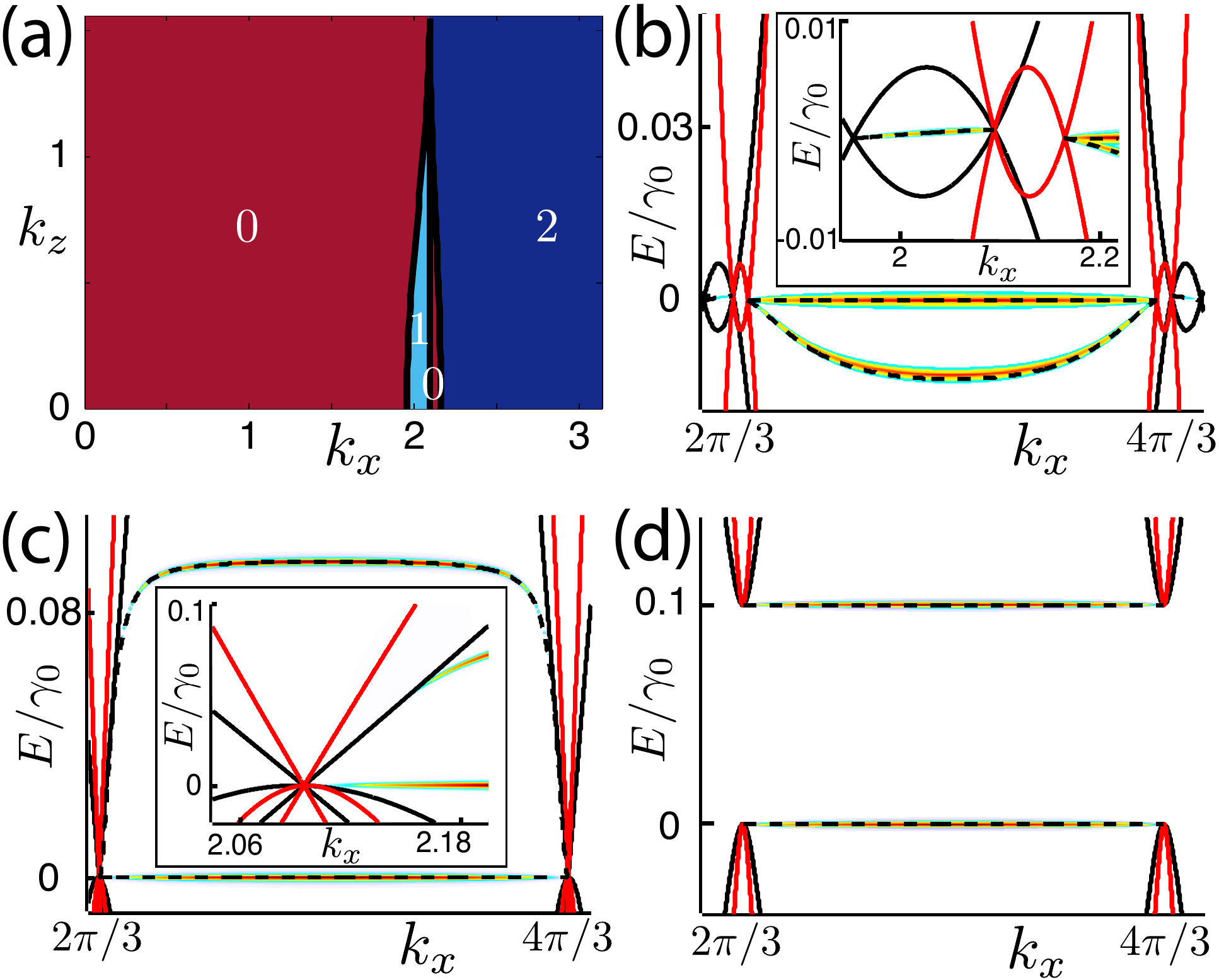}
  \caption{\label{fig:dispersions}  (a) $W(k_x,k_z)$ in the presence of the chiral symmetry ($\Delta=\gamma_4=0$). The transitions between different $W$ occur at $k_x=K_x-q_c(k_z)$, $k_x=K_x$ and $k_x=K_x+q_c(k_z)/2$, where the bulk energy gap closes. The regions $k_x \in [-\pi,0]$ or $k_z \in [-\pi/2,0]$ are mirror images of (a). (b)-(d) Surface state dispersions as a function of $k_x$ for (b) $\Gamma(k_z)=1$, (c) $\Gamma(k_z)= \Delta/(2 \gamma_1)$ and (d) $\Gamma(k_z)=0$ ($\Delta,\gamma_4 \neq 0$). In Figs.~(b), (c) the dashed black lines show the analytic approximations given by Eqs.~(\ref{E1E2}) and (\ref{E3}). In Fig.~(d) the dashed black lines show the exact surface state dispersions $E=0$ and $E=\Delta$ ($2\pi/3 \leq k_x \leq 4 \pi/3$) for Hamiltonian (\ref{full-H}). The colorful lines show the numerically computed surface state dispersions obtained with the help of surface Green functions (see App.~\ref{AppD}).  The transition between the two qualitatively different types of spectrums [(b) and (d)] occurs in the vicinity of the nexus (c). The solid lines (black and red) show the bulk dispersions along the special directions defined by the mirror planes. 
}
\end{figure}

We consider a translationally invariant system in the $x$- and $z$-directions, so that $k_x$ and $k_z$ are good quantum numbers. In the presence of chiral symmetry ($\Delta=\gamma_4=0$) the 1D Hamiltonian $H_{k_x, k_z}(k_y)$ has a well-defined topological invariant (winding number) $W(k_x, k_z)$ whenever there is no gap closing as a function of $k_y$ (see App.~\ref{sec:supp_chiral_mirror}), and this invariant determines the number of zero-energy surface states for each $k_x$ and $k_z$.
Moreover, $W(k_x, k_z)$ can change only at the Dirac lines with  $k_x=K_x-q_c(k_z)$, $k_x=K_x$ and $k_x=K_x+q_c(k_z)/2$, where the energy gap closes \cite{footnote}.  
By computing $W(k_x, k_z)$ in the presence of chiral symmetry (see App.~\ref{sec:supp_chiral_mirror}) we arrive at a flat band (zero energy) spectrum in the regions of the transverse momenta with $W\ne 0$ in Fig.~\ref{fig:dispersions}(a). There are two flat bands connecting the projected Dirac lines $k_x=K_x+q_c(k_z)/2$ and $k_x=K_x'-q_c(k_z)/2$. Additionally there is a single flat band between the projected Dirac lines at  $k_x=K_x-q_c(k_z)$ and $k_x=K_x$ \cite{footnote}.

In the presence of the chiral symmetry the Dirac lines meet at the boundary of the Brillouin zone $k_z=\pi/2$ and there is no three-fold degenerate exceptional point where a gap between the electron and hole bands opens up. Therefore, $\Delta$ and $\gamma_4$ have important effects on the surface state spectrum. The eigenstates in the presence of chiral symmetry can be solved exactly, and the surface state dispersions can be computed perturbatively in the limit $\gamma_0, \gamma_1 \Gamma(k_z)  \gg  \gamma_3 \Gamma(k_z), \gamma_4\Gamma(k_z), \Delta$. By utilizing also the observation that the surface bands connect the Dirac lines in energy and momentum we get  (see App.~\ref{AppC}) 
\begin{eqnarray}
E_1(k_x,k_z) &\approx&  4 \gamma_3^2 \Gamma^2(k_z)  \left[\Delta \gamma_0 -8 \gamma_1 \gamma_4 \Gamma ^2(k_z) \right]/\gamma_0^3, \nonumber \ \\ 
E_2(k_x, k_z) &\approx&   \frac{- 8 \frac{\gamma_1 \gamma_4}{\gamma_0} \Gamma^2(k_z)   +  \Delta }{ 1+\frac{4 \gamma_1^2  \Gamma^2(k_z) }{\gamma_0^2(1-4\cos^2(k_x/2))^2}} , 
\label{E1E2}
\end{eqnarray}
for $K_x+q_c(k_z)/2<k_x<K_x'-q_c(k_z)/2$ and 
\begin{eqnarray}
E_3 (k_x, k_z) & \approx & \frac{E_c(k_z)}{4}\bigg( 1- \sqrt{1-8\frac{k_x-K_x}{q_c(k_z)}} \bigg)^2
\label{E3}
\end{eqnarray}
for $K_x-q_c(k_z)<k_x<K_x$, which are valid far away from the nexus \cite{footnoted} (Fig.~\ref{fig:dispersions}). Therefore for small $k_z$ below the nexus the dispersions take the form of a drumhead that is bounded by the projected crossing points of the electron- and hole-like bands \cite{Chiu14,Chan15} [Fig.~\ref{fig:dispersions}(b)].
On the other hand, for large $k_z$ on the other side of the nexus [$\Gamma(k_z)< \Delta/(2 \gamma_1)$] there is a gap between electron- and hole-like bands and one of the surface bands connects two electron-like bands to each other, and the other surface band connects two hole-like bands. For $k_z=\pi/2$  the surface state dispersions can be solved exactly for Hamiltonian (\ref{full-H}) and one obtains $E_1=0$ and $E_2=\Delta$ for $2\pi/3 \leq k_x \leq 4 \pi/3$ (see App.~\ref{AppC}) [Fig.~\ref{fig:dispersions}(d)].
This qualitative change in the behavior of the surface bands on opposite sides of the nexus signals the existence of an exceptional point in the momentum space where three bands are simultaneously degenerate. By numerically computing the surface Green functions \cite{surface-Green}, we find that in the vicinity of the nexus the surface states hybridize with the bulk states so that they connect bulk band edges to each other instead of being bounded by the projected Dirac lines [Fig.~\ref{fig:dispersions}(c)]. This hybridization appears in the regime where the Dirac cones around the band crossings [Eq.~(\ref{H0Hc})] are overtilted ($|d\xi_{c}/dq_x|>|d\vec{h}_{c}/dq_x|$) forming type II Dirac lines (see App.~\ref{AppA} and App.~\ref{AppD}) \cite{Volovik_typeII_main}.

\section{Summary and discussion} 
We have identified the symmetries of the model that allow stabilizing the nexus in the momentum space  and  shown how the momentum-space structure of surface states follows from the properties of Dirac lines. In the vicinity of the nexus the behavior of the surface states changes qualitatively, indicating the existence of triple degeneracy point in the momentum space. There is an ongoing search for new types of fermions in condensed matter systems (in addition to Majorana, Weyl and Dirac fermions) that are described by simultaneous crossings of multiple bands  \cite{Bradlyn16}. The best candidate material for the study of nexus fermions is regular  graphite, where the surfaces with a component parallel to the $c$-axis should exhibit surface states. The properties of these surface states can be studied with STM and ARPES. Other candidate materials include for example suitably stacked silicene layers \cite{silicene} and InAs$_{1-x}$Sb$_x$ \cite{Winkler16}. The latter also supports a pair of triple degeneracy points connected by Dirac lines. Moreover, this material obeys similar mirror and three-fold rotational symmetries as graphite, so that the stabilization of the nexus in InAs$_{1-x}$Sb$_x$ follows from our analysis. We point out that the symmetry analysis described in this paper may be a useful starting point for a general classification scheme of the nexus semimetals  based on the space group symmetries, and it would be interesting to find out whether the predicted properties of the surface state spectrum in the vicinity of the nexus are generic for all nexus semimetals.

Apart from spectroscopic features, topological phases often have unusual response characteristics (anomalies) 
\cite{Qi08, Ryu12, Dziom16, Nielsen83, Hosur-review, Volovik-review15}. One interesting direction for future research is to find out whether the nexus semimetal phase is characterized by an anomaly associated with a spectral flow between the nexus points. We also point out that
even relatively weak interactions can lead to symmetry-broken states at the surface because of the large density of states caused by the approximately flat bands \cite{Honerkamp00,Li13,Potter14, Schnyder-edge-instability, Yazyev10, Heikkila11b}. These symmetry-broken states are expected to be exotic since they cannot be described with a mean field theory \cite{Kauppila}. Finally, we expect that the interactions  may lead to  "dipole" correlations within the structure, so that the effective $\Delta$ is renormalized. This would mean that the position of the nexus and the size of the energy gap above it would depend on temperature.

\acknowledgements

We thank G.~E.~Volovik for fruitful discussions and G. Winkler for pointing us the reference \onlinecite{Winkler16}. This work was supported by the Academy of Finland Centre of Excellence program (project No. 284594) and the European Research Council (Grant No. 240362-Heattronics).

\appendix

\section{Detailed description of the symmetries of the model and the bulk properties \label{AppA}}

We consider the tight-binding model for Bernally stacked honeycomb lattices [Eq.~(\ref{full-H}) in the main text].
Similarly as in the main text we assume the hierarchy of couplings and consider all the couplings to be positive unless stated otherwise (see App. \ref{othersigns}).

\begin{figure*}
 \includegraphics[width=0.9\linewidth]{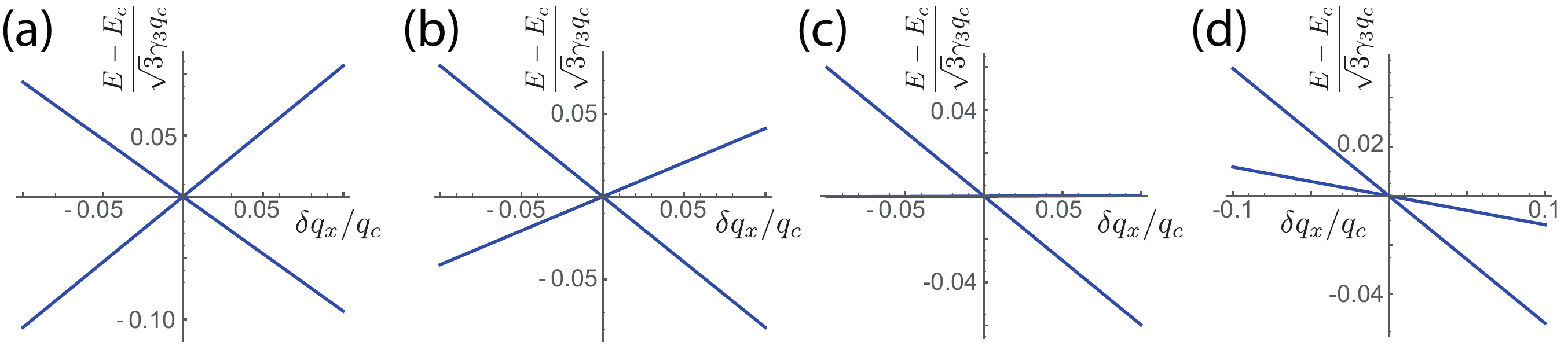}
  \caption{\label{fig:DiractypeII} Energy-momentum dispersions around the band crossing at $\mathbf{q}=q_c(-1, 0)$ for (a) $\Gamma(k_z)= 1$, (b) $\Gamma(k_z)= 0.6$, (c) $\Gamma(k_z)= 0.3$ and (d) $\Gamma(k_z)=0.2$. With increasing $k_z$ [decreasing $\Gamma(k_z)$] the Dirac cone becomes more tilted, and for sufficiently small $\Gamma(k_z)$ it is overtilted i.e. so-called type II Dirac cone. This transition occurs at slightly smaller value of $k_z$ [$\Gamma(k_z) \approx \Delta/\gamma_1$] than the nexus [$\Gamma(k_z) = \Delta/(2\gamma_1)$]. Tight-binding parameters are $\gamma_1=0.3 \gamma_0$, $\gamma_3=\Delta=0.1 \gamma_0$ and $\gamma_4=0.05 \gamma_0$.
}
\end{figure*}

The Hamiltonian obeys   a (i)  SU(2) spin rotation symmetry (block-diagonal in real spin), (ii) time-reversal symmetry $$H^*(-k_x,-k_y,-k_z)=H(k_x,k_y,k_z),$$ and (iii) several mirror symmetries
\begin{eqnarray*}
H(k_x, k_y, k_z)&=&H(-k_x, k_y, k_z), \\ 
H(k_x, k_y, k_z)&=&H(k_x, k_y, -k_z), \\ 
H(k_x, k_y, k_z)&=&\tau_x  \sigma_xH(k_x, -k_y, k_z)\tau_x  \sigma_x.
\end{eqnarray*}
Additionally, there exists (iv) a three-fold rotational symmetry
$$H(k_x, k_y, k_z)=H(\bar{k}_x, \bar{k}_y, k_z)=H(\tilde{k}_x, \tilde{k}_y, k_z),$$ 
where $\bar{k}_{x}=-k_x/2+\sqrt{3} k_y/2$, $\bar{k}_{y}=-k_y/2-\sqrt{3} k_x/2$, a $\tilde{k}_{x}=-k_x/2-\sqrt{3} k_y/2$ and $\tilde{k}_{y}=-k_y/2+\sqrt{3} k_x/2$, so that similar mirror symmetries exist also with respect to ($\bar{k}_{x}$, $\bar{k}_{y}$) and ($\tilde{k}_{x}$, $\tilde{k}_{y}$). In a special limit $\Delta=\gamma_4=0$ the system also supports a chiral symmetry $C H(\mathbf{k}) C=- H(\mathbf{k})$, where $C=\tau_0 \sigma_z$. 

The most important symmetries are the mirror symmetries with nontrivial matrix structure $H(k_x, k_y, k_z)=\tau_x  \sigma_xH(k_x, -k_y, k_z)\tau_x  \sigma_x$ [and correspondingly for ($\bar{k}_{x}$, $\bar{k}_{y}$) and ($\tilde{k}_{x}$, $\tilde{k}_{y}$)]. There are special planes going through the middle of the Brillouin zone and at the boundary of the Brillouin zone which are mapped  back to themselves in the mirror symmetries (up to a reciprocal lattice vector).  The relevant three planes around the $K$ point are directed along the $k_z$-direction and $k_y=2 \pi/\sqrt{3}$,  $\bar{k}_y=-2 \pi/\sqrt{3}$ and $\tilde{k}_y=0$ within the $(k_x, k_y)$-plane. 
Within these mirror planes the mirror symmetries give rise to  symmetries commuting with the Hamiltonian at fixed momentum
\begin{eqnarray*}
S^\dag H(k_x, 2 \pi/\sqrt{3} ,k_z) S&=&H(k_x, 2 \pi/\sqrt{3} ,k_z)\\
\bar{S}^\dag H(\bar{k}_x, -2 \pi/\sqrt{3}, k_z) \bar{S}&=&H(\bar{k}_x, -2 \pi/\sqrt{3}, k_z)\\
\tilde{S}^\dag H(\tilde{k}_x, 0, k_z) \tilde{S}&=&H(\tilde{k}_x, 0, k_z).
\end{eqnarray*}
The symmetry operators in different coordinates are 
 $S=U \tau_x \sigma_x U^\dag$, $\bar{S}=U^\dag \tau_x \sigma_x U$, $\tilde{S}=\tau_x \sigma_x$, where $U={\rm diag}(e^{-i 2\pi/3}, 1, e^{i 2\pi/3}, e^{-i 2\pi/3})$. All these symmetries are simultaneously valid within the line directed along $k_z$-direction at the $K$-point, which in different coordinates appears at  $K=(2 \pi/3, 2\pi/\sqrt{3})$, $\bar{K}=(2 \pi/3, -2\pi/\sqrt{3})$ and $\tilde{K}=(-4\pi/3, 0)$. 
 
 The Hamiltonian (\ref{full-H}) around the $K$-point can be expanded as  [$(\tilde{k}_{x}, \tilde{k}_{y})=\tilde{K}+(q_{x}, q_{y})$]
 \begin{widetext}
\begin{equation}
\tilde{H}= \begin{pmatrix}
  \Delta      & -  \frac{\sqrt{3}}{2} \gamma_0 (q_x-i q_y) &  \sqrt{3} \gamma_4 \Gamma(k_z)  (q_x+i q_y)      & -2 \gamma_1 \Gamma(k_z) \\
-  \frac{\sqrt{3}}{2} \gamma_0 (q_x+i q_y)    & 0 & \sqrt{3} \gamma_3 \Gamma(k_z)  (q_x-i q_y)     &  \sqrt{3} \gamma_4 \Gamma(k_z)  (q_x+i q_y)	   \\
 \sqrt{3} \gamma_4 \Gamma(k_z)  (q_x-i q_y)           & \sqrt{3} \gamma_3 \Gamma(k_z)  (q_x+i q_y)         & 0  & -  \frac{\sqrt{3}}{2} \gamma_0 (q_x-i q_y)  \\
   -2 \gamma_1 \Gamma(k_z)            & \sqrt{3} \gamma_4 \Gamma(k_z)  (q_x-i q_y)       & - \frac{\sqrt{3}}{2} \gamma_0 ( q_x+i q_y)  & \Delta  
\end{pmatrix}. 
\end{equation}
\end{widetext}
By analyzing this Hamiltonian it is easy to see that there always exists a band crossing at $(q_x, q_y)=(0,0)$ for all values of $k_z$ as discussed in the main text. Additionally there exists 
three other Dirac lines in the vicinity of $K$  point. These band crossings appear at finite energy $E_c$ within the three distinct mirror planes at  $(q_x, q_y)=q_c (-1,0)$ and $(q_x, q_y)=q_c (1/2, \pm \sqrt{3}/2)$, where
\begin{eqnarray}
E_c(k_z) &\approx&  \gamma_3^2 \left[4 \gamma_1^2 \Gamma^2 (k_z) -\Delta ^2\right] \nonumber \\ && \hspace{-0.5cm} \times \frac{\Delta \gamma_0^2 -8 \gamma_1 \gamma_4 \gamma_0 \Gamma^2(k_z)  + 4  \Delta  \gamma_4^2 \Gamma^2(k_z) }{\left[\gamma_1 \gamma_0^2 -2 \gamma_4 \Delta \gamma_0+4 \gamma_1 \gamma_4^2 \Gamma^2(k_z) \right]^2}, \label{Ec-supp}
\end{eqnarray}
\begin{equation}
q_c(k_z) \approx   \frac{2 \sqrt{3} \gamma_3 \Gamma(k_z) [4 \gamma_1^2 \Gamma(k_z)^2-\Delta^2] }{v_-^2 [2 \gamma_1 \Gamma(k_z)+\Delta]+v_+^2 [2 \gamma_1 \Gamma(k_z)-\Delta]}  \label{qc-supp}
\end{equation}
and $v_{\pm}^2= \frac{3}{4} [\gamma_0 \pm 2 \gamma_4 \Gamma(k_z)]^2$. Around the crossings at $\mathbf{q}=\mathbf{0}$ and  $\mathbf{q}=q_c(-1, 0)$  the projected Hamiltonians in the basis of eigenvectors corresponding to eigenvalues $+1$ and $-1$ of the mirror symmetry operator at the crossings are given by Eqs.~(\ref{H0Hc}) in the main text.
The Hamiltonian around $\mathbf{q}=q_c(-1, 0)$ describes a tilted anisotropic Dirac cone for each value of $k_z$. For small values of $k_z$ the Dirac cone is only slightly tilted but when one approaches the nexus [$\Gamma(k_z) =\Delta/(2\gamma_1)$] the tilt increases. The Dirac cone becomes overtilted at 
\begin{equation}
\Gamma(k_z^*) \approx \frac{\Delta}{\gamma_1}.
\end{equation}
For $k_z > k_z^*$, the Dirac line can thus be called type II Dirac line (see also the discussion in Ref.~\onlinecite{Volovik_typeII_main}).
This transition, which occurs already at slightly smaller value of $k_z$ than the nexus, is illustrated in Fig.~\ref{fig:DiractypeII}.

\section{The significance of the chiral and mirror symmetries: Flat bands and drumhead surface states \label{sec:supp_chiral_mirror}}

\begin{figure*}
 \includegraphics[width=0.78\linewidth]{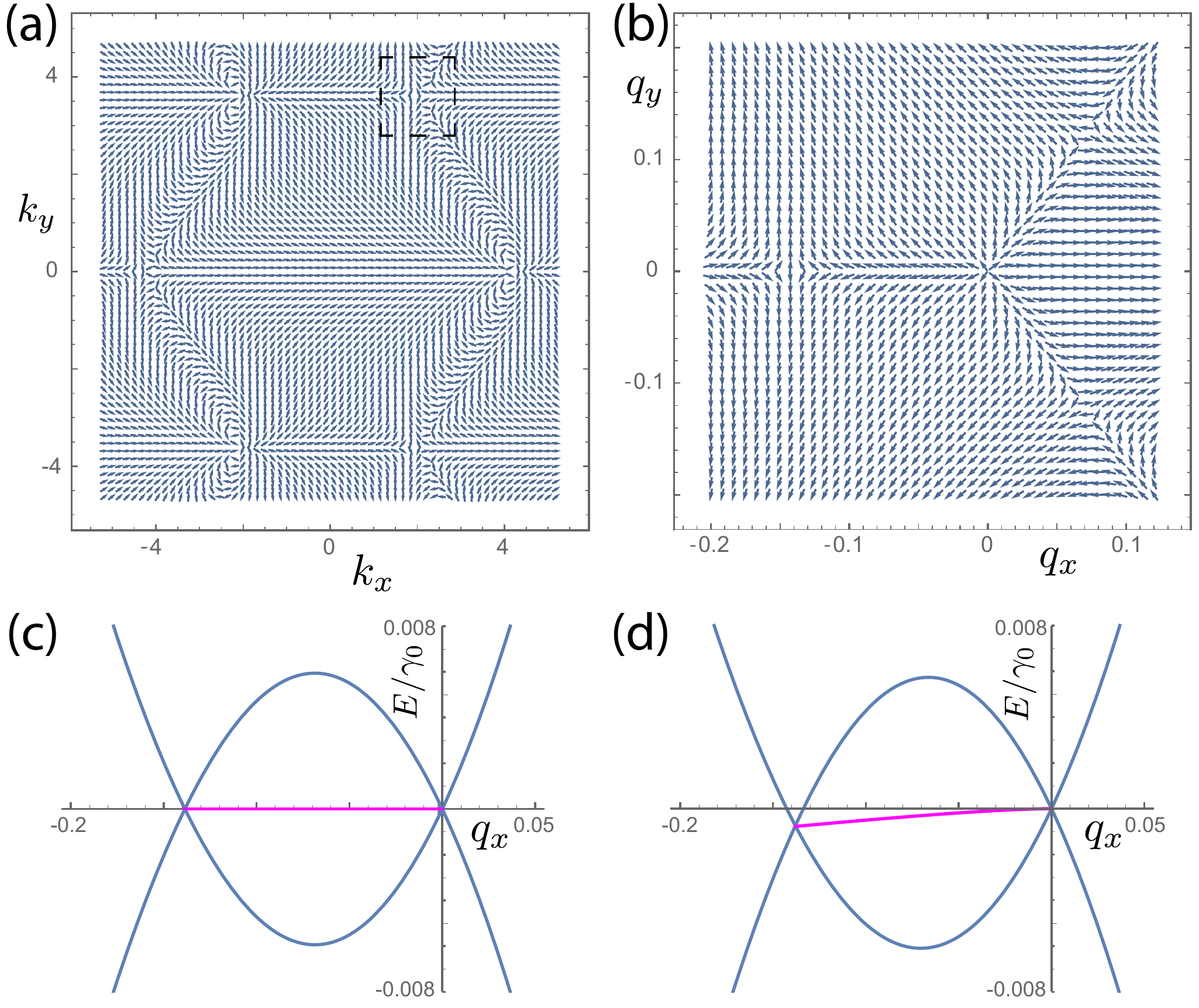}
  \caption{\label{fig:chiralvsmirror} (a) In the presence of chiral symmetry the topology of the system is described by complex field $z(\mathbf{k})$ [Eq.~(\ref{W-supp})], which is represented here with arrows. The complex field $z$ has vortex/antivortex lines at the positions of the Dirac lines. (b) Magnification of (a) around the $K$-point (the boxed region). (c) In the presence of chiral symmetry the surface states (magenta) form flat bands at $E=0$ between the Dirac lines. The number of flat bands is determined by the winding number $W(k_x,k_z)$. (d) In absence of chiral symmetry the band crossings (protected by mirror symmetry) can appear at finite energy $E_c(k_z)$. In the regime $K_x-q_c(k_z)< |k_x|<K_x$ there exists a protected drumhead surface state (magenta) bounded by the projected Dirac lines. In the regime $K_x+q_c(k_z)/2< |k_x|<\pi$ the existence of two drumhead surface states depends on how strongly the Dirac cone is tilted (see below). The parameters used in the figures are $\gamma_1=0.3 \gamma_0$, $\gamma_3=0.1 \gamma_0$ and $\Gamma(k_z)=1$. In (a)-(c) we have assumed chiral symmetry $\Delta=\gamma_4=0$. In (d) $\Delta=0.1 \gamma_0$ and $\gamma_4=0.05 \gamma_0$.
}
\end{figure*}

We consider a translationally invariant system in the $x$- and $z$-directions corresponding to a zigzag edge for each graphene layer. Therefore $k_x$ and $k_z$ are good quantum numbers and by fixing them we get a 1D Hamiltonian $H_{k_x, k_z}(k_y)$, which depends only on $k_y$. In the special limit  $\Delta=\gamma_4=0$,  the system supports an additional chiral symmetry $C=\tau_0 \sigma_z$, and the 1D Hamiltonians $H_{k_x, k_z}(k_y)$ have well-defined topological invariants. To calculate this topological invariant, we first notice that the Hamiltonian can be written in a block-off-diagonal form
\begin{equation}
U^\dag H(\mathbf{k}) U=\begin{pmatrix}
0 & A(\mathbf{k}) \\
A^\dag (\mathbf{k}) & 0
\end{pmatrix}, \label{Block-off}
\end{equation}
where
\begin{eqnarray}
A(\mathbf{k})= \begin{pmatrix}
- \gamma_0  f_1(k_x,k_y)  & -2 \gamma_1 \Gamma(k_z) \\
 2 \gamma_3 \Gamma(k_z)  f_2^*(k_x,k_y)     &  - \gamma_0 f_1(k_x,k_y)	   			
\end{pmatrix}, \label{A-supp}
\end{eqnarray} 
$f_1(k_x,k_y)=e^{- i \vec{\delta} \cdot (k_x,k_y)} f(k_x,k_y)$, $f_2(k_x,k_y)=e^{2 i \vec{\delta} \cdot (k_x,k_y)}  f(k_x,k_y)$ and  $\vec{\delta}=(0,1/(2\sqrt{3}))$.
The topological invariant can then be defined as a winding number
\begin{eqnarray}
W(k_x, k_z)=\frac{i}{2 \pi} \int \frac{dz(k_y)}{z}, \ z=\frac{\det[A(\mathbf{k})]}{|\det[A(\mathbf{k})]|},
 \label{W-supp}
\end{eqnarray} 
where the integration is over the 1D Brillouin zone in $k_y$ direction. 

The winding number undergoes a series of transitions at the momenta of the projected Dirac lines $k_x=K_x-q_c(k_z)$, $k_x=K_x$ and $k_x=K_x+q_c(k_z)/2$  in such a way that 
\begin{equation}
W(k_x,k_z) = \begin{cases}
    0, &  |k_x|<K_x-q_c(k_z)\\
     1,  & K_x-q_c(k_z)< |k_x|<K_x \\
     0, &  K_x< |k_x|<K_x+q_c(k_z)/2\\
     2, & K_x+q_c(k_z)/2< |k_x|<\pi. \\
\end{cases} \label{Wregimes}
\end{equation}
These changes occur because the complex field $z$ has vortex lines at the positions of the Dirac lines [Fig.~\ref{fig:chiralvsmirror}(a),(b)].   The winding number $W$ as a function of $k_x$ and $k_z$ is shown in Fig.~\ref{fig:dispersions}(a) in the main text.  

The winding number $W(k_x, k_z)$ determines the number of zero-energy states for each $k_x$ and $k_z$. In the presence of chiral symmetry, the band crossings always occur at energy $E=0$, and the surface states form flat bands at $E=0$ between the Dirac lines [Fig.~\ref{fig:chiralvsmirror}(c)].

In the absence of chiral symmetry the band crossings are protected by the mirror symmetry and they appear at finite energy 
$E_c$. Moreover around some of the crossings the low energy theory is described by a tilted Dirac cone. By considering a general tilted Dirac cone
\begin{equation}
H_{T}(q_x, q_y)=(E_c+b_0 q_x) \sigma_0 +  b_1 q_x \sigma_z +b_2 q_y \sigma_y,
\end{equation}
we can introduce an edge by replacing $q_y = -i\partial_y$. By looking for an exponentially localized solution 
$$\psi=\begin{pmatrix}
 a_1\\a_2  \end{pmatrix} e^{-\alpha y}$$ 
at the energy $E_s(q_x)$, we get
\begin{eqnarray}
\alpha^2=\frac{b_1^2q_x^2 -(b_0 q_x+E_c-E_s)^2}{b_2^2}, \nonumber\\ (b_0 q_x+E_c-E_s+b_1 q_x) a_1=-\alpha b_2 a_2.
\end{eqnarray}
Therefore real solutions of $\alpha$ exist only if 
\begin{equation}
E_c+b_0q_x-b_1q_x \leq E_s(q_x) \leq E_c+b_0q_x+b_1q_x \label{inequalities}
\end{equation}
and by varying $E_s(q_x)$ (and the sign of $q_x$) one can interpolate between different boundary conditions determining the ratio $a_2/a_1$. [The surface state may occur either for $q_x<0$ or $q_x>0$ depending on the boundary conditions. Similarly the exact dispersion $E_s(q_x)$ depends on the boundary conditions.] As one can see from inequalities (\ref{inequalities}), for $q_x=0$ the surface state energy must satisfy $E_s(0)=E_c$, so that a single Dirac line gives rise to a drumhead surface state dispersion bounded by the projected Dirac line in energy and momentum. 
In particular, it follows from this calculation that the existence of a single surface state is always guaranteed in the regime $K_x-q_c(k_z)< |k_x|<K_x$ [Fig.~\ref{fig:chiralvsmirror}(d)].  On the other hand, in the regime $K_x+q_c(k_z)/2< |k_x|<\pi$ we expect to find two surface states in the presence of chiral symmetry, and once the chiral symmetry is broken due to $\Delta, \gamma_4 \ne 0$ the existence of the drumhead surface states may depend on how strongly the Dirac cones are tilted (see below).

\section{Analytical results for the surface state spectrum \label{AppC}}

In order to obtain analytical insights into the surface state dispersions, we start by considering the zero energy wave functions in the presence of the chiral symmetry ($\gamma_4=\Delta=0$). The  solutions exist either only in sublattice A or sublattice B depending on which surface one is considering. In the following we concentrate on those solutions which are localized in sublattice B
\begin{equation}
\psi(y)=\begin{pmatrix}
\psi_{B1}(y)  \\ \psi_{B2}(y)   
\end{pmatrix} e^{i K_y y}, \label{ansatz}
\end{equation}
where $K_y=2\pi/\sqrt{3}$ is the $y$-component of the momentum at the Dirac point and the indices refer to the layer degree of freedom. By  substituting the ansatz [Eq.~(\ref{ansatz})] to the block off-diagonal form of the Hamiltonian [Eq.~(\ref{Block-off})], we arrive at equations 
\begin{eqnarray}
-\gamma_0 \hat{F}_1(k_x,-i\partial_y) \psi_{B1}(y)&=&2\gamma_1 \Gamma(k_z) \psi_{B2}(y)  \nonumber \\
-2 \gamma_3 \Gamma(k_z)\hat{F}_2(k_x,-i\partial_y) \psi_{B1}(y)  &=& \gamma_0 \hat{F}_1(k_x,-i\partial_y) \psi_{B2}(y), \nonumber
\end{eqnarray}
where
\begin{eqnarray*}
F_1(k_x, q)&=& 2 \cos(k_x/2)- e^{-i \sqrt{3}q/2}, \nonumber \\
F_2(k_x, q)&=& e^{-i \sqrt{3}q/2} \bigg[2 \cos(k_x/2)- e^{i \sqrt{3}q/2} \bigg].
\end{eqnarray*}

We look for a solution of the form (the plane $y=y_0$ describes the surface)
\begin{equation}
\psi_{B1}(y)= b_1 e^{-Q (y-y_0)}, \ \psi_{B2}(y)= b_2 e^{-Q (y-y_0)},
\end{equation}
which gives 
\begin{equation}
\det \begin{pmatrix}
- \gamma_0 F(k_x,iQ)  & -2 \gamma_1 \Gamma(k_z) \\
 -2 \gamma_3 \Gamma(k_z)   F_2(k_x,iQ)     &  - \gamma_0 F(k_x,iQ) 	   			
\end{pmatrix}=0. \label{det-eq}
\end{equation}
The wave functions localized in sublattice B exist on the right surface, i.e., ${\rm Re}[Q]<0$. For $K_x-q_c<k_x<K_x$ we obtain
\begin{eqnarray}
Q &\approx& \frac{-q_c}{2} \bigg(-1+\sqrt{1-8\frac{k_x-K_x}{q_c}}+ 2\frac{k_x-K_x}{q_c}\bigg), \nonumber \\ \frac{b_2}{b_1} &\approx& \sqrt{3} \frac{ \gamma_0}{\gamma_1} \frac{Q+k_x-K_x}{4 \Gamma(k_z)}.
\end{eqnarray}
This solution describes how the weight of the wave function within the different layers varies as a function of $k_x$ and $k_z$ in the case of the flat band corresponding to $W=1$ regime in Eq.~(\ref{Wregimes}).

Additionally we need to find expressions for the wave functions of the flat bands correponding to  $W=2$ regime. To describe these wave functions we look for solutions of Eq.~(\ref{det-eq}) allowing also complex values of $Q$, but still requiring ${\rm Re} [Q]<0$. Such solutions exist for $K_x+q_c/2<k_x<\pi$ and they come in pairs $Q$ and $Q^*$, where 
\begin{widetext}
\begin{equation}
Q=\frac{2}{\sqrt{3}}\ln\bigg\{2(1+\frac{\sqrt{3}}{4} q_c)\cos(k_x/2)+i\sqrt{-\frac{\sqrt{3}}{2} q_c\bigg[1+\frac{\sqrt{3}}{4} q_c+2(1+\frac{\sqrt{3}}{8} q_c)\cos k_x\bigg]} \bigg\}.
\end{equation}
\end{widetext}
The orthonormal solutions obtained using these solutions can be written as 
\begin{eqnarray}
\psi_1(y)&=&\frac{1}{{\cal N}_1}\begin{pmatrix}
   1 \\ b_2  
\end{pmatrix}  e^{-Q (y-y_0)} e^{i K_y y},  \nonumber \\ \psi_2(y)&=&\frac{1}{{\cal N}_2} \bigg[\begin{pmatrix}  1 \\ b_2^*  
\end{pmatrix}  e^{-Q^* (y-y_0)}+A\begin{pmatrix}
  1 \\ b_2  
\end{pmatrix}  e^{-Q (y-y_0)} \bigg] e^{i K_y y}, \nonumber \\ \label{exact-sol-gen-orth}
\end{eqnarray}
where
\begin{eqnarray}
 b_2&=&-\frac{\gamma_0 F(k_x,iQ)}{2 \gamma_1 \Gamma(k_z)}, \hspace{0.3cm} A=-\frac{1+(b_2^*)^2}{1+|b_2|^2}\frac{1-e^{-\sqrt{3} |{\rm Re}[Q]|}}{1-e^{\sqrt{3} Q^*}}, \nonumber \\ {\cal N}_1&=&\sqrt{\frac{1+|b_2|^2}{1-e^{-\sqrt{3} |{\rm Re}[Q]|}}}, \nonumber \\ {\cal N}_2&=&\sqrt{\frac{(1+|b_2|^2)(1+|A|^2)}{1-e^{-\sqrt{3} |{\rm Re}[Q]|}}+2 \ {\rm Re}\bigg[\frac{A^*(1+b_2^{*2})}{1-e^{\sqrt{3} Q^*}}\bigg]}. \label{discrete-coeff}
\end{eqnarray}
In the special limit $\Gamma(k_z)=0$, the Hamiltonian becomes block-diagonal in the layer degree of freedom, so that $Q=Q^*$ and these expressions describe two copies of the edge states for a single layer graphene. On the other hand for $\Gamma(k_z)=1$ the solutions describe the edge states of bilayer graphene.

By considering the couplings $\gamma_4$ and $\Delta$ as a perturbation, we obtain in lowest order in $\gamma_3$ 
\begin{eqnarray}
E_1 &\approx& 0 +{\cal O}(\gamma_3^2), \nonumber \\ E_2 &\approx&   \frac{- 8 \frac{\gamma_1 \gamma_4}{\gamma_0} \Gamma^2(k_z)   +  \Delta }{ \bigg( 1+\frac{4 \gamma_1^2  \Gamma^2(k_z) }{\gamma_0^2(1-4\cos^2(k_x/2))^2}\bigg)} + {\cal O}(\gamma_3^2) 
\end{eqnarray}
for $K_x+q_c(k_z)/2<|k_x|<\pi$ and 
\begin{equation}
E_3 \approx  \bigg[\Delta -\frac{8 \gamma_1 \gamma_4}{\gamma_0} \Gamma^2  \bigg] \frac{ \gamma_3^2 \Gamma^2}{\gamma_0^2}  \bigg( 1- \sqrt{1-8\frac{k_x-K_x}{q_c(k_z)}} \bigg)^2
\end{equation}
for $K_x-q_c(k_z)<k_x<K_x$.

Additionally, we can utilize the knowledge that far away from the nexus the surface states must have the shape of the drumhead bounded by the projected Dirac lines. By including the necessarily corrections to $E_1$ and $E_2$, we arrive to surface states dispersions described by Eqs.~(\ref{E1E2})  in the main text.
In principle these expressions should work for $\Gamma(k_z) \gg \Delta/2\gamma_1$ or $|k_x-K_x| \gg \Delta/\gamma_0$. However, numerically we find that they work everywhere except very close to the nexus. On the other hand, since the single protected surface state should exist also in the vicinity of the nexus, we find that the surface dispersion $E_3$ is given by Eq.~(\ref{E3}) in the main text.

Finally we notice that the full tight-binding Hamiltonian also has an exact solution for $\Gamma(k_z)=0$ i.e. for $k_z=\pi/2$. Namely, in this case the Hamiltonian becomes block-diagonal in the layer degree of freedom, and each block can be solved similarly as the surface states for a single layer graphene. This way, we obtain 
\begin{equation}
E_1=0, \ E_2=\Delta \   \textrm{for } \Gamma(k_z)=0 \textrm{ and } K_x \leq |k_x| \leq \pi. \label{exact-sol-supp}
\end{equation}
These also coincide with Eqs.~\eqref{E1E2} for $\Gamma(k_z)=0$.

\section{Numerical analysis of the surface state spectrum in the vicinity of the nexus \label{AppD}}

\begin{figure*}
  \includegraphics[width=0.68\linewidth]{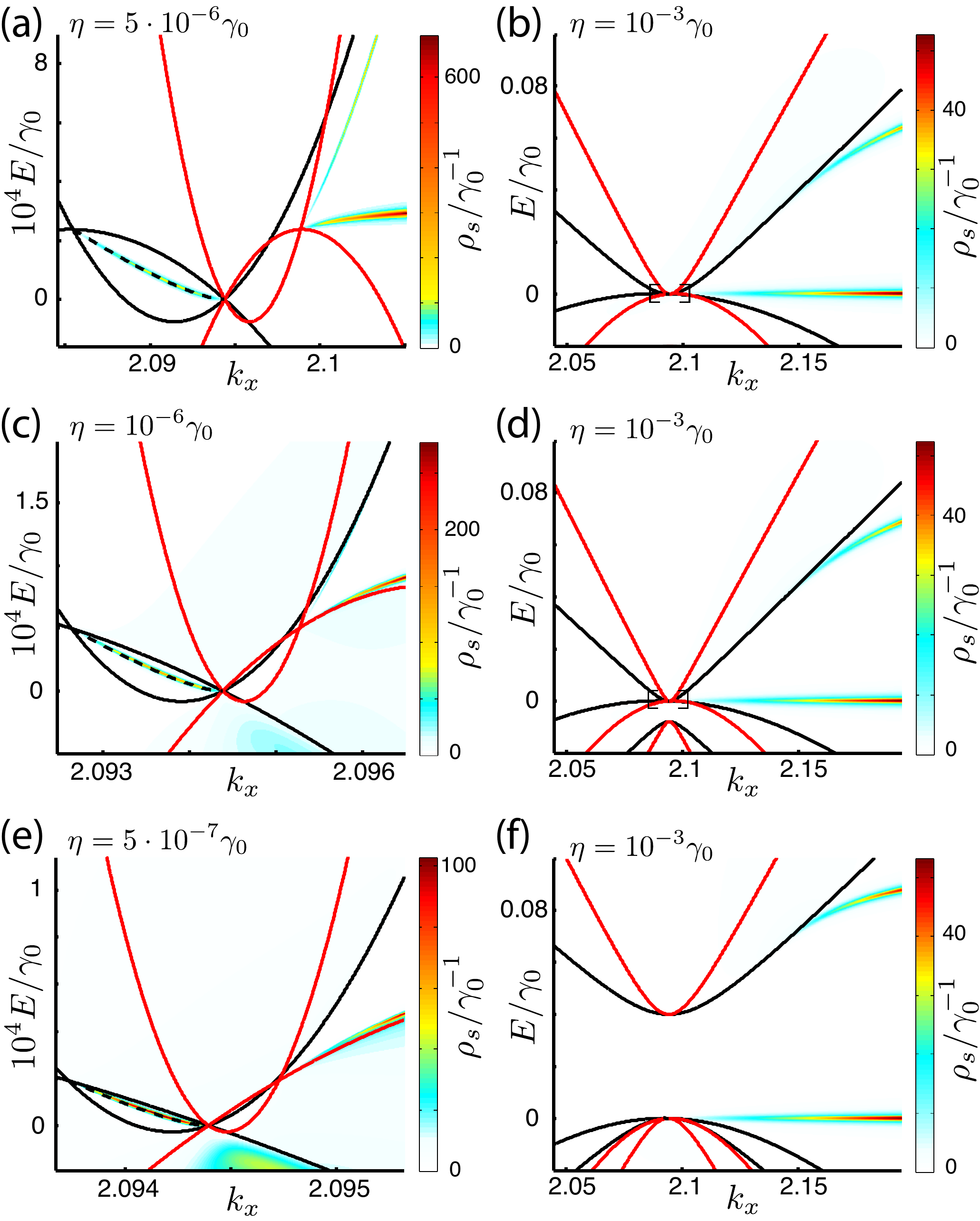}
  \caption{\label{fig:surfacestatessupp}  Surface state dispersions as a function of $k_x$ for (a) $\Gamma(k_z)=0.3$, (b),(c) $\Gamma(k_z)= 0.2$, (d),(e) $\Gamma(k_z)=0.18$ and (f) $\Gamma(k_z)=0.1$. Figures (c) and (e) show the spectrum around the boxed region in (b) and (d), respectively. The dashed black lines show the analytic approximation given by Eq.~(\ref{E3}), which works very well also in the vicinity of the nexus. The colorful lines show the numerically computed surface state dispersions obtained by plotting the surface density of states as a contour plot. (The broadening $\eta$ is chosen in each figure in such a way that the surface state dispersions are well visible in the pictures.)   The solid lines (black and red) show the bulk dispersions along the special directions defined by the mirror planes. For small $k_z$ sufficiently far from the nexus the surface state dispersions take a form of a drumhead that is bounded by the projected crossing points of the electron- and hole-like bands [figure (a)]. When the nexus is approached by decreasing  $\Gamma(k_z)$ the bulk Dirac cones become strongly overtilted and the surface states hybridize with bulk states [figures (b)-(e)]. On the other side of the nexus [$\Gamma(k_z)< \Delta/(2 \gamma_1)$] there is a gap between electron- and hole-like bands and one of the surface bands connects two electron-like bands to each other, and the other surface band connects two hole-like bands [figure (f)]. With increasing $k_z$ these surface state dispersions smoothly deform towards the exact analytic solution for $k_z=\pi/2$ [Eq.~(\ref{exact-sol-supp})]. Tight-binding parameters are $\gamma_1=0.3 \gamma_0$, $\gamma_3=\Delta=0.1 \gamma_0$ and $\gamma_4=0.05 \gamma_0$.
}
\end{figure*}

We have checked that the analytical solutions (\ref{E1E2}), (\ref{E3}) and (\ref{exact-sol-supp}) describe the surface state dispersions reasonably far away from the nexus by numerically diagonalizing the tight-binding Hamiltonian in the case of a finite width in $y$-direction. However, in the vicinity of the nexus it is difficult to obtain analytic expressions for the surface state dispersions. Moreover, in that regime the localization length of the surface states becomes very long, and thus the numerical diagonalization of the tight-binding Hamiltonian with large enough width in $y$-direction also becomes computationally expensive. 

Alternatively the surface state dispersions can be obtained by numerically calculating the surface Green function $G_{s}^R(E, k_x, k_z)=[E+i\eta - H(k_x, k_z) ]^{-1}_{00}$ (the matrix indices $00$ correspond to the surface in the $y$-direction). The surface density of states is given by
\begin{equation}
\rho_s(E, k_x, k_z)=-\frac{1}{\pi} {\rm Im}[{\rm Tr} \ G_{s}^R(E, k_x, k_z)].
\end{equation}
This method is computationally much more efficient since the Green function can be computed for a semi-infinite system using a quickly converging renormalization group method\cite{surface-Green}. In the numerics we broaden the $\delta$-peaks in the $\rho_s(E, k_x, k_z)$ corresponding to the surface state energies $E_s(k_x, k_z)$ by using a nonzero value of $\eta$. 

Sufficiently far away from the nexus, the Green function method reproduces the analytic surface state dispersions (\ref{E1E2}), (\ref{E3}) and (\ref{exact-sol-supp})  as shown in Fig.~\ref{fig:dispersions}(b) and (d) in the main text. For small $k_z$ the dispersions take a form of a drumhead that is bounded by the projected crossing points of the electron- and hole-like bands [Fig.~\ref{fig:dispersions}(b) in the main text], but for $k_z=\pi/2$ one of the surface bands connects two electron-like bands to each other, and the other surface band connects two hole-like bands [Fig.~\ref{fig:dispersions}(d) in the main text]. We now turn to the description of the transition between these qualitatively distinct regimes, which occurs in the vicinity of the nexus.

Figure \ref{fig:surfacestatessupp} shows the surface density of states $\rho_s(E, k_x, k_z)$ as a function of $k_x$ and $E$ for specific values of $k_z$. The surface state energies $E_s(k_x, k_z)$ show up as peaks in $\rho_s(E, k_x, k_z)$ (broadened by $\eta$) and form lines as a function of $k_x$. Sufficiently far from the nexus the surface state dispersions take a form of a drumhead that is bounded by the projected crossing points of the electron- and hole-like bands [Fig.~\ref{fig:surfacestatessupp}(a)]. When the nexus is approached by decreasing $\Gamma(k_z)$  [$\Gamma(k_z) \to \Delta/(2\gamma_1)]$ the bulk Dirac cones become strongly overtilted and the surface states hybridize with bulk states so that they connect bulk band edges to each other instead of being bounded by the projected Dirac lines [Fig.~\ref{fig:surfacestatessupp}(b)-(e)]. On the other side of the nexus [$\Gamma(k_z)< \Delta/(2 \gamma_1)$] one of the surface bands connects two electron-like bands to each other, and the other surface band connects two hole-like bands [Fig.~\ref{fig:surfacestatessupp}(f)]. With increasing $k_z$ these dispersions approach the exact solution for $k_z=\pi/2$ [Eq.~(\ref{exact-sol-supp})].

\begin{figure*}
 \includegraphics[width=0.8\linewidth]{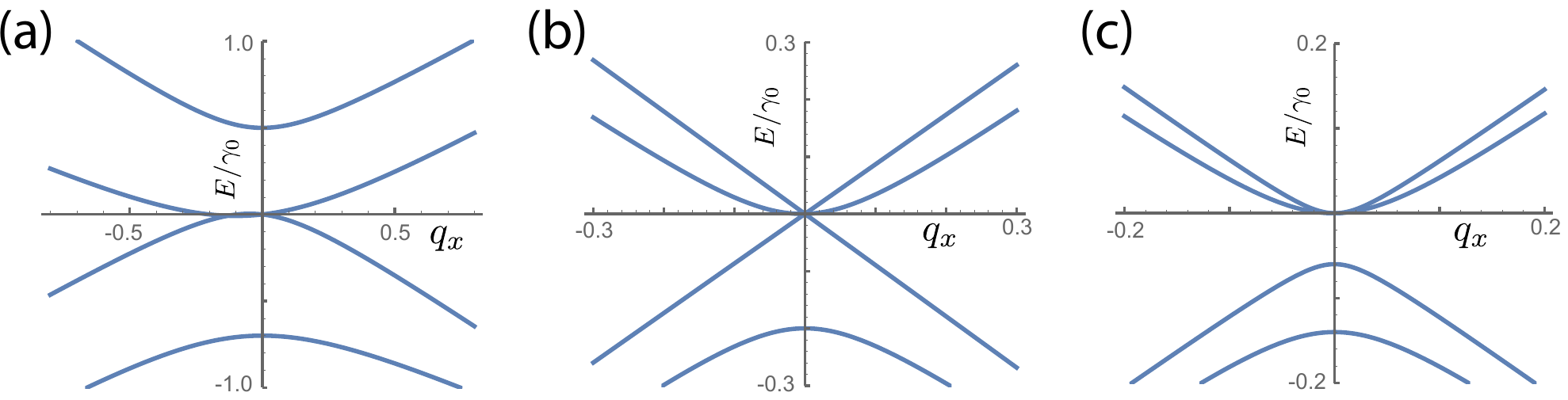}
  \caption{\label{fig:nexussupp} Energy-momentum dispersions for bulk bands around the $K$ point for $\Delta<0$ and (a) $\Gamma(k_z)= 1$, (b) $\Gamma(k_z)= |\Delta|/(2 \gamma_1)$ and (c) $\Gamma(k_z)= |\Delta|/(2 \gamma_1)-0.1$. The figures look similar to the case $\Delta>0$ (cf.~Fig.~\ref{fig:nexus} in the main text) except that one needs to mirror all bands in energy $E \to -E$. In particular for large $k_z$ [$\Gamma(k_z) < |\Delta|/(2 \gamma_1)$] two electron (hole) bands are degenerate  at the $K$ point for $\Delta<0$  ($\Delta>0$). Tight-binding parameters are $\gamma_1=0.3 \gamma_0$, $\gamma_3=0.1 \gamma_0$, $\Delta=-0.1 \gamma_0$ and $\gamma_4=0.05 \gamma_0$.
}
\end{figure*}

\begin{figure*}
  \includegraphics[width=1.4\columnwidth]{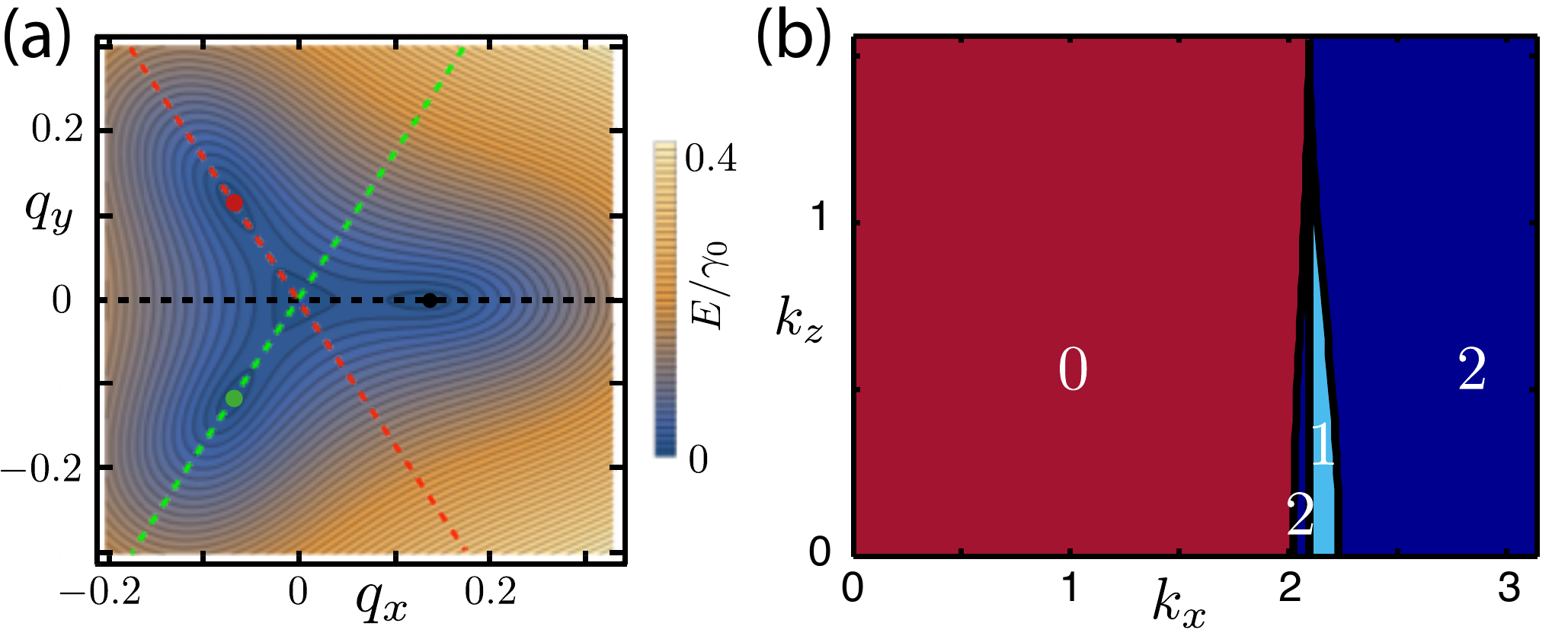}
  \caption{\label{fig:minusgamma} (a) Momentum-dependent energy gap  $E_3(\mathbf{k})-E_2(\mathbf{k})$ around the $K$ point  for $k_z=0$ and $\gamma_3<0$.  Similarly, as in the case $\gamma_3>0$ [Fig.~\ref{fig:Diraclines}(a) in the main text] there exists four band crossing points but now they are located in the momentum space on the opposite side of the $K$ point.  (b)  Winding number $W(k_x,k_z)$ in the presence of the chiral symmetry ($\Delta=\gamma_4=0$) for $\gamma_3<0$. The transitions between different $W$ occur at $k_x=K_x-|q_c(k_z)|/2$, $k_x=K_x$ and $k_x=K_x+|q_c(k_z)|$, where the bulk energy gap closes. Because the bulk band crossings are on the opposite side of the $K$ point, the fine structure of $W(k_x, k_z)$ around $k_x \approx K_x$ is modified in comparison to the case $\gamma_3>0$ [cf.~Fig.~\ref{fig:dispersions}(a) in the main text]. Tight-binding parameters are $\gamma_1=0.3 \gamma_0$, $\gamma_3=-0.1 \gamma_0$, $\Delta=0.1 \gamma_0$ and $\gamma_4=0.05 \gamma_0$.}
\end{figure*}

\section{Effects of the different signs of tight-binding parameters \label{othersigns}}

The signs of $\gamma_0$ and $\gamma_1$ can be chosen arbitrarily  in the Hamiltonian (\ref{full-H}) without loss of generality, so we choose $\gamma_0, \gamma_1>0$. The relative signs of the other parameters then influence the physics.

\begin{figure*}
  \includegraphics[width=0.68\linewidth]{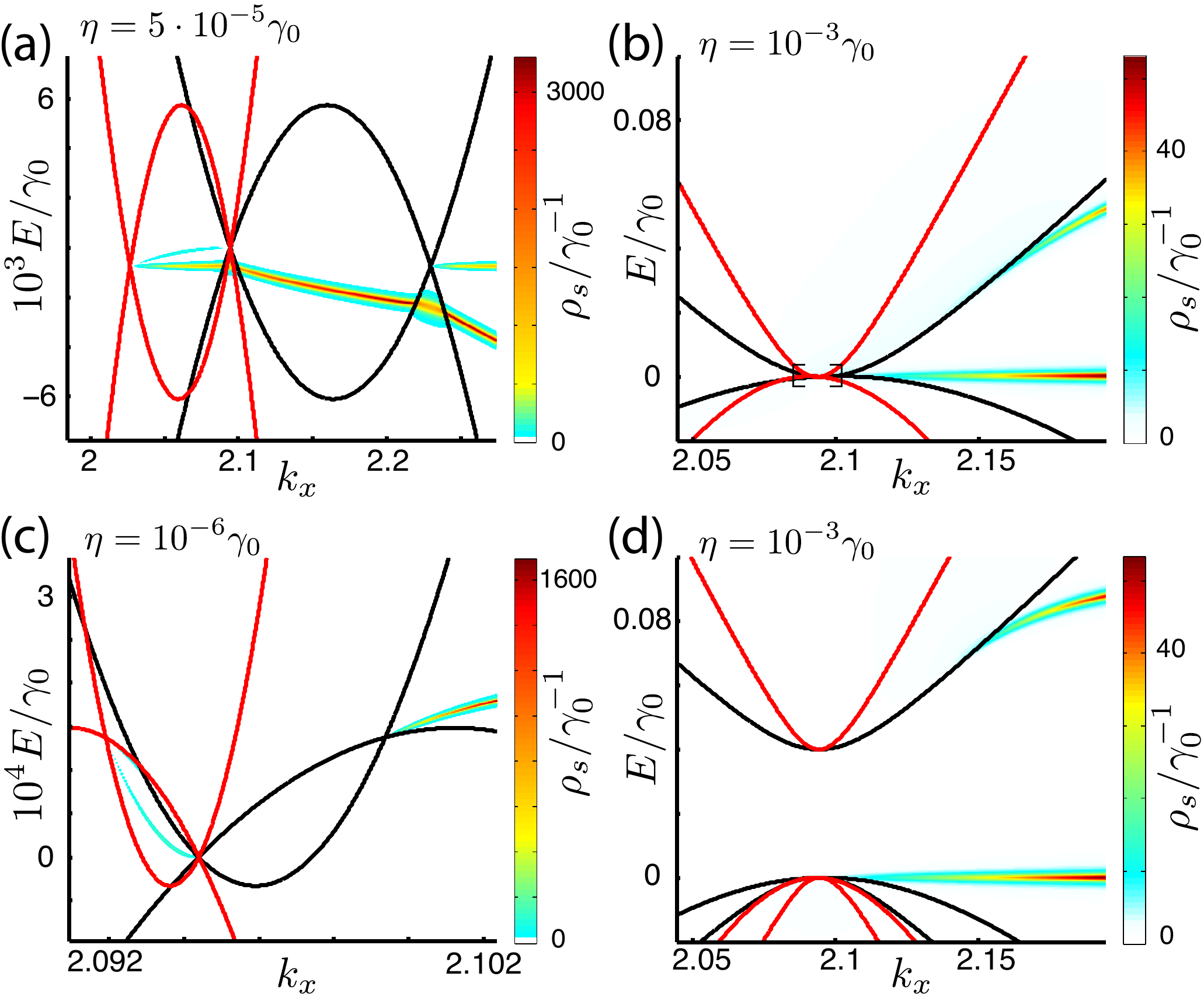}
  \caption{\label{fig:surfacestatessupp2}  Same as Fig.~\ref{fig:surfacestatessupp} but for $\gamma_3=-0.1 \gamma_0$ and (a) $\Gamma(k_z)=1$, (b),(c) $\Gamma(k_z)= 0.25$, (d) $\Gamma(k_z)=0.1$. Figure (c) shows the spectrum around the boxed region in (b). For small $k_z$ sufficiently far from the nexus [figure (a)] two drumhead surface states originate from $k_x=K_x-|q_c(k_z)|/2$. One of them connects to the Dirac line at $k_x=K_x$ and other one to a Dirac line at $k_x=K_x'+|q_c(k_z)|/2$. (The spectrum around $K'$ point is obtained by mirroring the spectrum around the $K$ point.) Additionally there is a drumhead surface state connecting Dirac lines at $k_x=K_x+|q_c(k_z)|$ and $k_x=K_x'-|q_c(k_z)|$. When the nexus is approached by decreasing  $\Gamma(k_z)$ the bulk Dirac cones become overtilted and the surface states hybridize with bulk states [figures (b)-(c)]. On the other side of the nexus [$\Gamma(k_z)< \Delta/(2 \gamma_1)$] the spectrum is similar as in the case  $\gamma_3>0$.
}
\end{figure*}

The sign of $\Delta$ influences the nature of the bands involved in the transition occurring at the nexus $\Gamma(k_z)= |\Delta|/(2 \gamma_1)$. Namely for $\Delta>0$ the transition occurs as illustrated in Fig.~\ref{fig:nexus} in the main text. For small $k_z$ 
electron and hole bands are degenerate at the $K$ point, whereas for large $k_z$  
two hole bands are degenerate at the $K$ point and there is an energy gap between the electron and hole bands. In the case $\Delta<0$, for small $k_z$ 
electron and hole bands are still degenerate at the $K$ point [Fig.~\ref{fig:nexussupp}(a)]. However, now two electron bands are degenerate at the $K$ point for large $k_z$ [Fig.~\ref{fig:nexussupp}(c)]. In general all the results remain qualitatively similar when the sign of $\Delta$ is changed except that one needs to mirror all the figures in energy $E \to -E$.

The sign of $\gamma_4$  mainly influences the band crossing energies $E_c(k_z)$ as can be seen from Eq.~(\ref{Ec-supp}). Since the drumhead surface states connect band crossings in energy and momentum, $\gamma_4$ also influences the surface state dispersions [Eqs.~(\ref{E1E2}) and (\ref{E3})]. In particular the relative sign of $\gamma_4$ and $\Delta$ determines whether the drumhead dispersion in the regime $K_x-q_c(k_z)<|k_x|<K_x$ is tilted in the same direction for all values of $k_z$ (unidirectional surface states). This can for example influence the transport properties of these systems.

The sign of $\gamma_3$ determines the side in which the Dirac lines are with respect to the $K$-point [cf.~Fig.~\ref{fig:Diraclines}(a) in the main text and Fig.~\ref{fig:minusgamma}(a)]. As a result the winding number $W(k_x, k_z)$ in the presence of chiral symmetry ($\Delta=\gamma_4=0$) is modified in the different regions of $k_x$ and $k_z$ around $k_x \approx K_x$ [cf.~Fig.~\ref{fig:dispersions}(a) in the main text and Fig.~\ref{fig:minusgamma}(b)]. As discussed in App.~\ref{sec:supp_chiral_mirror}, the winding number influences the number of surface states also when the chiral symmetry is broken ($\Delta, \gamma_4 \ne 0$). We expect that for small $k_z$ far away from the nexus there exists drumhead surface states bounded by the projected Dirac lines, and the number of these surface states for each $k_x$ and $k_z$ is determined by $W(k_x, k_z)$.  This expectation is confirmed by the surface state spectrum shown in Fig.~\ref{fig:surfacestatessupp2}(a). On the other hand, we find that close to the nexus the surface states hybridize with bulk states similarly as in the case $\gamma_3>0$ [Fig.~\ref{fig:surfacestatessupp2}(b),(c)].  On the other side of the nexus [$\Gamma(k_z)< \Delta/(2 \gamma_1)$] one of the surface bands connects two electron-like bands to each other, and the other surface band connects two hole-like bands, so that the surface state spectrum is practically indistinguishable from the surface state spectrum in the case $\gamma_3>0$ [cf.~Fig.~\ref{fig:surfacestatessupp2}(d) and Fig.~\ref{fig:surfacestatessupp}(f)].


\begin{thebibliography}{99}
\bibitem{footnote}
Similar results are also valid around the $K'$ point but we focus on the $K$ point here.
\bibitem{footnotec}
The three-fold degenerate point  where a gap between the electron and hole bands opens up is necessarily a meeting point of the Dirac lines associated with the band crossings between the electron and hole bands. However, the opposite is not necessarily true. It is possible to construct Hamiltonians where the Dirac lines have additional meeting points without the three-fold degeneracy of the bands.
\bibitem{footnoteb}
The Hamiltonians are unique up to an unimportant multiplication of the off-diagonal elements with $e^{i \phi}$ and $e^{-i \phi}$.
\bibitem{footnoted}
In principle these approximations should work for $\Gamma(k_z) \gg \Delta/2\gamma_1$ or $|k_x-K_x| \gg \Delta/\gamma_0$. However, in practice they work everywhere except very close to the nexus.
\end{thebibliography}
\end{document}